\documentstyle[mprocl,psfig]{article}

\def\Journal#1#2#3#4{{#1} {\bf #2}, #3 (#4)}

\def\PRL{\em Phys. Rev. Lett.}
\def\PRD{{\em Phys. Rev.} D}

\def\lap{\hbox{${_{\displaystyle<}\atop^{\displaystyle\sim}}$}}
\def\gap{\hbox{${_{\displaystyle>}\atop^{\displaystyle\sim}}$}}
\def\be{\begin{equation}}
\def\ee{\end{equation}}
\def\bea{\begin{eqnarray}}
\def\eea{\end{eqnarray}}
\def\etal{{\it et al}.}
\begin{document}

\title{A CRITICAL LOOK AT MASSIVE SCALAR FIELD COLLAPSE}

\author{C.M. CHAMBERS}

\address{Department of Physics, 
         Montana State University, 
	 Bozeman,
         MT 59717-3840, USA}

\author{P.R. BRADY}

\address{Theoretical Astrophysics, 
         Mail-Code 130-33, 
	 California Institute of Technology, 
	 Pasadena, CA 91125, USA}

\author{S.M.C.V. GON\c{C}ALVES}

\address{Department of Physics,
         University of Newcastle upon Tyne, 
         NE1 7RU, UK}

\maketitle

\abstracts{We present the findings of an investigation of critical
behavior in the collapse of spherically symmetric distributions
of massive scalar field. Two distinct types of phase transition
are observed at the verge of black hole formation and a
criterion for determining when each type of transition will
occur is given.}

%
%%%%%%%%%%%%%%%%%%%%%%%%%%%%%%%%%%%%%%%%%%%%%%%%%%
%          MAIN TEXT SECTION                     %
%%%%%%%%%%%%%%%%%%%%%%%%%%%%%%%%%%%%%%%%%%%%%%%%%%
%
\section{Introduction}
\label{cmc-s10}

Critical behavior in gravitational collapse is a fascinating area of
research within classical General Relativity and exemplifies the role
played by non-linear dynamics at the verge of black hole
formation. Since Choptuik's discovery of critical point behavior in
the collapse of spherically symmetric distributions of real, massless
scalar field,\cite{cmc-b10} similar phenomenology has been observed 
in other models of gravitational collapse.  Here we summarize 
the results of an investigation of critical behavior in the 
collapse of massive spherically symmetric distributions of real scalar
field,\cite{cmc-b20} described by the Einstein-Klein-Gordon (EKG)
system of equations. Its attractiveness, as another model 
exhibiting critical point behavior, is further enhanced by 
the two following properties:
	\begin{enumerate} 
	  \item 
            The characteristic length associated with the mass
            $\mu$ destroys the scale invariance of the EKG 
	    equations.
          \item 
	    The EKG system admits unstable soliton-like 
	    solutions.\cite{cmc-b30}
          \end{enumerate} 
These observations suggest that the qualitative
picture of critical point behavior in massive scalar field collapse
could differ from the massless limit and might be similar to that
found by Choptuik
\etal~\cite{cmc-b40} in their study of the collapse of 
a Yang-Mills field. 

%%%%%%%%%%%%%%%%%%%%%%%%%%%%%%%%%%%%%%%%%%%%%%%%%%%%%%
\section{Results}
\label{cmc-s20}
%%%%%%%%%%%%%%%%%%%%%%%%%%%%%%%%%%%%%%%%%%%%%%%%%%%%%%

\begin{figure}[ht]
	\centerline{\hbox{
	\psfig{figure=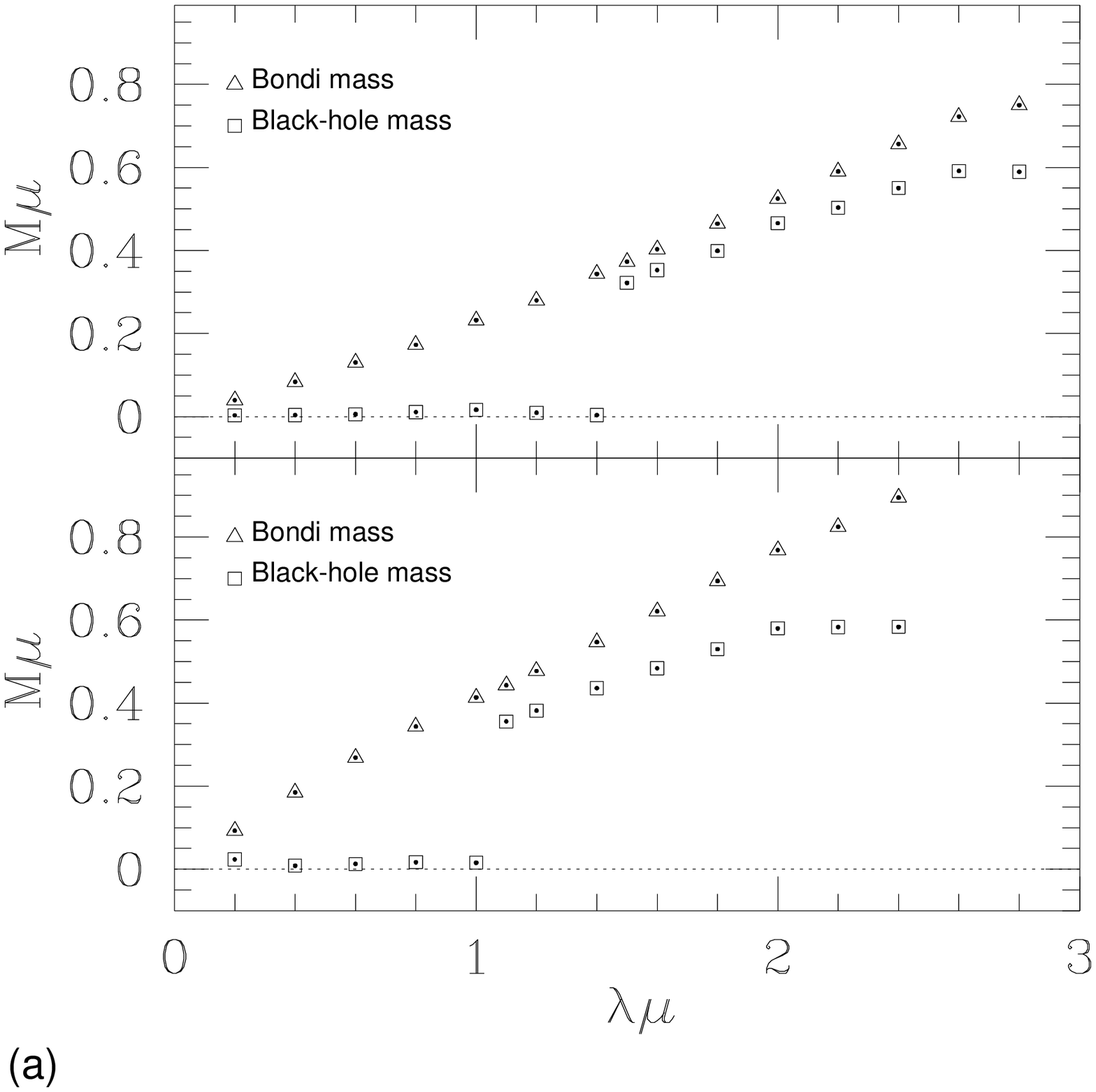,height=2.5in}
	\psfig{figure=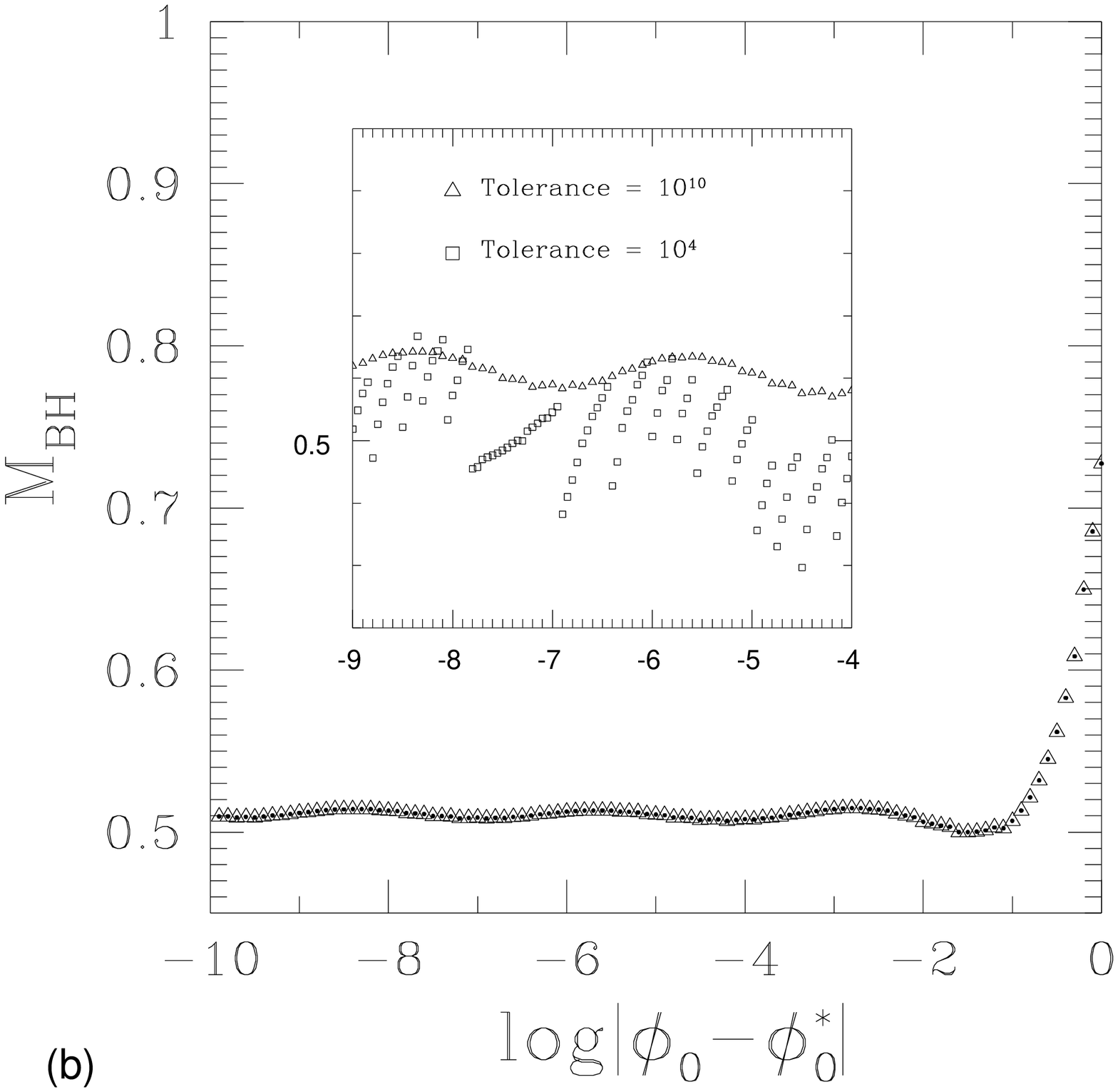,height=2.5in}
	}}
	\caption{(a) The Bondi mass of the initial field
	pulse, and the black hole mass at criticality,
	versus the radial extent of the field, for the initial
	data sets (i) and (ii) of Table~\ref{cmc-t10}.
	(b)  The black hole mass spectrum for supercritical
	Type I evolutions. The inset demonstrates the spurious
	discontinuities that can occur if tolerances are not
	set high enough.}
        \label{cmc-f10}
	\end{figure}

We find two distinct types of phase transition occur in 
the collapse of a massive scalar field: 
	\begin{tabbing}
         {\it Type I } \=  : \=
         Black hole formation turns on at finite mass and the 
         critical solutions are \\ \> \> unstable soliton stars 
	 with masses $M_{\rm s} \lap 0.6 \mu^{-1}$.  \\
         {\it Type II} \> : \>  
         Black hole formation turns on at infinitesimal mass and 
         the critical soluti- \\
	 \> \> on is identical to that found by Choptuik 
         in the collapse of massless scalar \\
	 \> \> fields. 
	\end{tabbing}
We also formulate a criterion for determining
when each type of phase transition will occur,  and which helps
to clarify the role that intrinsic scales play in critical collapse.
If $\lambda$ is the radial extent of the field in its initial
configuration,\footnote{For generic initial data it is not possible
to define the radial extent of the field. However, for initial data
sets (i) and (ii) in Table~\ref{cmc-t10} we define
$\lambda = 2 \sigma$.}  then
	\begin{itemize}
    	  \item Type I behavior occurs when $\lambda \mu \gap 1$,
	  \item Type II behavior occurs when $\lambda \mu \lap 1$.
      \end{itemize}
Intuitively, this selection rule
seems reasonable since one
expects the massive results to differ from the massless 
results~\cite{cmc-b10} only if the Compton wavelength of the field
$\mu^{-1}$ is smaller than the radial extent of the initial field
pulse. Indeed, Fig.~\ref{cmc-f10}~(a) shows the Bondi mass of the initial
field pulses, described by data sets (i) and (ii) of 
Table~\ref{cmc-t10}, and the resulting black hole masses,
at the critical point $\phi_{0} = \phi_{0}^{*}$. 
Type I behavior is clearly evident when $\lambda \mu \gap 1$,
whilst Type II behavior is observed for $\lambda \mu \lap 1$.
Figure~\ref{cmc-f10}~(a) also indicates that
the interface between Type I and Type II phase transitions
occurs when the Bondi mass of the initial pulse,
$M_{\rm B} \sim 0.4 \mu^{-1}$.
%
%%%%%%%%%%%%%%%%%%%%%%%%%%%%%%%
%       FIGURE 1              %
%%%%%%%%%%%%%%%%%%%%%%%%%%%%%%%
%

Figure~\ref{cmc-f10}~(b) shows the black hole mass spectrum,
$M_{\rm BH}$ vs.~$\log | \phi_{0} -\phi_{0}^{*} |$, for 
supercritical, Type I, evolutions with $\mu =1$. 
The results displayed are for initial data set (iii) of 
Table~\ref{cmc-t10}. Here we can see the mass gap at the
threshold of black hole formation is $M_{\rm gap} \sim 0.51 
\mu^{-1}$.  In general the mass gap lies in the range
$0.35 \lap \mu M_{\rm gap} \lap 0.59$, with the upper limit
being fixed by the maximum mass a soliton star can 
have.\cite{cmc-b30} The inset of Fig.~\ref{cmc-f10}~(b)
demonstrates the care that must be taken when setting
the tolerance that decides if, and when, a black hole
forms. For Type II behavior, a tolerance of $10^{4}$ 
is sufficient to achieve accurate results. However,
for Type I transitions, if the tolerance is too
low, we observe spurious discontinuities in the 
mass spectrum.\footnote{An explanation of the 
tolerance and the discontinuities in the black
hole mass spectrum is given by Brady \etal\cite{cmc-b20}}
The inset shows the mass spectrums for black hole formation 
tolerances of $10^{4}$ and $10^{10}$, obtained under 
identical evolutions. While a tolerance of $10^{4}$
exhibits widespread discontinuities, raising the 
tolerance to $10^{10}$ reveals them to be purely numerical
effects.  This withstanding, the oscillation
imposed on the mass spectrum is not an artifact of the
numerics but is similar to the fine structure found
by Hod and Piran~\cite{cmc-b50} in the massless results.
%
%%%%%%%%%%%%%%%%%%%%%%%%%%%%%%%
%       TABLE  1              %
%%%%%%%%%%%%%%%%%%%%%%%%%%%%%%%
%
        \begin{table}[ht]
	\caption{Three typical initial data sets considered in the
	collapse of  a massive scalar field. The parameters of the set,
	which may be varied, are shown under {\it Parameters} while the
	types of phase transition which may occur are shown
	under {\it Type}. \label{cmc-t10}}
	\vspace{0.4cm}
	\begin{center}
	\begin{tabular}{|c|l|c|c|}
	\hline
	Set  & \multicolumn{1}{c|}{$\phi(u=0,r)$}  &  Parameters   &  Type \\
	\hline
        (i)  &  $\phi_{0}r^{2}\exp [-(r-r_{0})^2/\sigma^{2}]$  &
	$\sigma$,~$\phi_{0}$  & I,~II  \\
	(ii) &  $\phi_{0}(1-\tanh [(r-r_{0})/\sigma])$  &
	$\sigma$,~$\phi_{0}$  & I,~II  \\
	(iii)  &  $\phi_{0} r (r+r_{0})^{-\sigma}/(1+\exp [r])$  &
	$\sigma$,~$\phi_{0}$  & I,~II \\
	\hline
	\end{tabular}
	\end{center}
	\end{table}
\section{Conclusions}
We find the presence of an intrinsic length scale 
changes the nature of critical phenomenon in
the collapse of a scalar field and speculate
that unstable, confined solutions could act
as critical solutions in other matter models. 
%
%%%%%%%%%%%%%%%%%%%%%%%%%%%%%%%%%%%%%%%%%%%%%%%%%%
%          ACKNOWLEDGMENTS                       %
%%%%%%%%%%%%%%%%%%%%%%%%%%%%%%%%%%%%%%%%%%%%%%%%%%
%
\section*{Acknowledgments}
CMC is a Fellow of the Royal Commission for the
Exhibition of 1851 and gratefully acknowledges
their financial support. PRB  is supported in
part by NSF Grant No. AST-9417371, and by a PMA 
Division Prize Fellowship at Caltech. SMCVG is
supported by the Programa PRAXIS XXI of the 
J.N.I.C.T of Portugal.
%
%%%%%%%%%%%%%%%%%%%%%%%%%%%%%%%%%%%%%%%%%%%%%%%%%%
%          BIBLIOGRAPHY                          %
%%%%%%%%%%%%%%%%%%%%%%%%%%%%%%%%%%%%%%%%%%%%%%%%%%
%

\end{document}